\documentclass[reprint,superscriptaddress,nofootinbib,amsmath,amssymb,aps]{revtex4-2}
\usepackage{graphicx}
\usepackage{dcolumn}
\usepackage{bm}
\usepackage{hyperref}
\hypersetup{colorlinks,citecolor=blue,linkcolor=blue,urlcolor=blue,}

\begin{document}

\title{Construction of fuzzy dark matter halos with arbitrary initial velocities}
\author{Yu-Ming Yang}
\email{yangyuming@ihep.ac.cn}
 \affiliation{%
 Key Laboratory of Particle Astrophysics, Institute of High Energy Physics, Chinese Academy of Sciences, Beijing 100049, China}
\affiliation{
 University of Chinese Academy of Sciences, Beijing 100049, China 
}%

\author{Xiao-Jun Bi}
\email{bixj@ihep.ac.cn}
\affiliation{%
 Key Laboratory of Particle Astrophysics, Institute of High Energy Physics, Chinese Academy of Sciences, Beijing 100049, China}
\affiliation{
 University of Chinese Academy of Sciences, Beijing 100049, China 
}%
\author{Peng-Fei Yin}
\email{yinpf@ihep.ac.cn}
\affiliation{%
 Key Laboratory of Particle Astrophysics, Institute of High Energy Physics, Chinese Academy of Sciences, Beijing 100049, China}

\begin{abstract}
Cosmological simulations of fuzzy dark matter (FDM) are computationally expensive, and the resulting halos lack flexibility in parameter adjustments, such as virial mass, density profile, and global velocity. Previous studies have introduced a method for constructing FDM halos with predefined density profiles. In this study, we investigate the initial global velocity of these constructed halos and find that it is nonzero. We provide the theoretical formula for this velocity and illustrate that it arises from the interference between states of odd and even parity. Our calculated results closely match simulation outcomes. Additionally, we showcase how to counteract this velocity and create a halo with a customizable initial global velocity. Our study presents a practical method for adjusting the initial global velocity of halos in controlled FDM simulations, facilitating investigations into tidal effects, galaxy collisions, and other scenarios.
\end{abstract}


\maketitle
\section{Introduction}
Fuzzy dark matter (FDM) \cite{hu2000fuzzy,peebles2000fluid,hui2017ultralight,hui2021wave,Ferreira_2021}, also known as ultralight bosonic dark matter or wave dark matter, stands out as a highly appealing candidate for dark matter. FDM is composed of ultralight bosons with a mass of approximately $10^{-22}$eV, resulting in a de Broglie wavelength of $\mathcal{O}(1)$ kpc at typical galaxy velocities. The large de Broglie wavelength imparts distinctive characteristics of FDM on galactic scales, potentially resolving the challenges encountered by cold dark matter (CDM) \cite{klypin1999missing,de2010core,boylan2011too,tulin2018dark} at small scales while maintaining consistency with CDM predictions at large scales.

Previous cosmological simulations have achieved significant success in revealing the properties of FDM. It has been noted that a FDM halo consists of a solitonic core along with a Navarro-Frenk-White-(NFW) like envelope \cite{Schive_2014_1,Schive_2014,Mocz_2017,Veltmaat_2018,zimmermann2024dwarf}. The soliton represents the ground state solution of the equation of motion governing FDM, which can be described as a Schr$\ddot{\text{o}}$dinger-Poisson system. The NFW-like envelope is composed of the excited state of this equation. The interference between the ground state and the excited states gives rise to soliton oscillations and a random walk phenomenon in the vicinity of the halo's central region \cite{Schive_2020,Li_2021}. Furthermore, the interference among the excited states leads to evolving granular structures within the NFW-like envelope \cite{Liu_2023}.

Nevertheless, current cosmological simulations of FDM also have some limitations \cite{Lin:2018whl}. One of the key challenges is the computational complexity involved in generating halos of substantial mass while maintaining spatial resolution below the de Broglie wavelength. Moreover, it is challenging for cosmological simulations to precisely produce desired halos with specific properties like virial mass, density profile, or global velocity that can be manually prescribed. This limitation becomes particularly relevant in controlled simulations for galaxy studies, such as those involving dynamical heating effect \cite{Bar_Or_2019,El_Zant_2019,Dutta_Chowdhury_2021,Dutta_Chowdhury_2023,Yang:2024vgw,Yang}, tidal stripping effect \cite{Du:2018qor,Hertzberg:2022vhk,Widmark:2023dec}, and galaxy collisions \cite{Paredes:2015wga,Schwabe:2016rze}, where halos with adjustable parameters are essential. To address this issue, Lin et al. \cite{Lin:2018whl} and Yavetz et al. \cite{Yavetz:2021pbc} have proposed methods for constructing FDM halos through eigenstate decomposition. These approaches enable the construction of a halo based on a relatively arbitrary input density profile.

In this study, we identify the presence of an initial global velocity when evolving the wave function constructed based on the initial density profile, which serves solely as an initial condition. This phenomenon is attributed to the introduction of phases in the wave function that are not constrained by the initial density profile. We provide the formula for this initial global velocity and ascertain that our theoretical predictions closely align with the results of simulations.  In the scenario of an isolated halo without initial velocity, adjustments to the phases in the initial wave function are necessary to ensure a total initial momentum of zero. However, this approach is difficult, in practice, due to the large number of phases associated with eigenstates. As an alternative, we demonstrate the efficacy of applying a Galilean boost to counteract the undesired motion of the halo. Conversely, through this approach, we can generate a halo with a customizable initial velocity.


The paper is organized as follows. In Sec. \ref{Sec_2},  we introduce the halos constructed in this study and highlight the nonzero initial global velocity observed in simulations. In Sec. \ref{Sec_3}, we provide an intuitive understanding and present the formula for the initial global velocity, along with a comparative analysis between our theoretical predictions and simulation results. In Sec. \ref{Sec_4}, we illustrate the method for offsetting the undesired velocity or creating a halo with a customizable initial global velocity. Finally, the conclusions of our study are summarized in  Sec. \ref{Sec_5}.

\section{Simulation method\label{Sec_2}}
The evolution of the FDM wave function is described by the Schr$\ddot{\text{o}}$dinger-Poisson (SP) equations \cite{hui2021wave}
\begin{equation}
    \begin{aligned}
        i\hbar \partial_t \psi&=-\frac{\hbar^2}{2m_a}\nabla^2\psi+m_a\Phi\psi,\\
        \nabla^2\Phi&=4\pi G\rho, \quad \rho=m_a|\psi|^2.
    \end{aligned}
    \label{SPequ}
\end{equation}
In cosmological simulations, it has been observed that a FDM halo consists of a solitonic core and a NFW-like envelope \cite{Schive_2014_1,Schive_2014,Mocz_2017,Veltmaat_2018}. In this study, we construct three halos with different FDM masses of $m_a/10^{-23}\text{ eV}=1,~3$, and $5$, respectively. We do not consider a FDM particle mass on the order of 
$10^{-22}$ eV due to the increasing computational demands associated with higher FDM particle masses. This heightened computational complexity arises from the need for constructing the initial wave function with more eigenstates, as well as the requirement for higher resolution and shorter time steps during the simulation process. The target FDM profile we utilize is given by \textcolor{red}{\cite{Bernal_2017,Meinert:2021gpb,Chan:2021bja,Liu_2023}}
\begin{equation}
    \rho_\text{in}(r)=\left\{\begin{aligned}
        &\frac{\rho_c}{\left[1+0.091(r/r_c)^2\right]^8},\quad r<kr_c\\
        &\frac{\rho_s}{(r/r_s)\left(1+r/r_s\right)^2}e^{-r^2/2r_\text{cut}^2},\quad r\geq kr_c.
        \end{aligned}\right.
\label{target}
\end{equation}
The profile within $r<kr_c$ corresponds to a soliton profile that aligns well with simulation results  \cite{Schive_2014_1,Schive_2014}. The outer region comprises a NFW-like envelope \cite{Navarro_1997}, which is modulated by an exponential factor to mitigate boundary effects in simulations. In this equation, the parameters $k,~r_s$, and $r_\text{cut}$ and the halo mass within $r_\text{cut}$ are set to be $3,~10~$kpc, $50~$kpc and $1\times 10^{10}M_\odot$, respectively.
The solitonic core density and core radius are connected by the scaling relation \cite{Guzman_2006,Dutta_Chowdhury_2023}
\begin{equation}
    \rho_c=1.95\times 10^7 M_\odot \text{kpc}^{-3}\left(\frac{m_a}{10^{-22}\text{ eV}}\right)^{-2}\left(\frac{r_c}{\text{kpc}}\right)^{-4}.
    \label{scaling}
\end{equation}
Thus, the target halo profile can be uniquely determined by the scaling relation and the continuity condition at $kr_c$.

\begin{figure}
  \includegraphics[width=\linewidth]{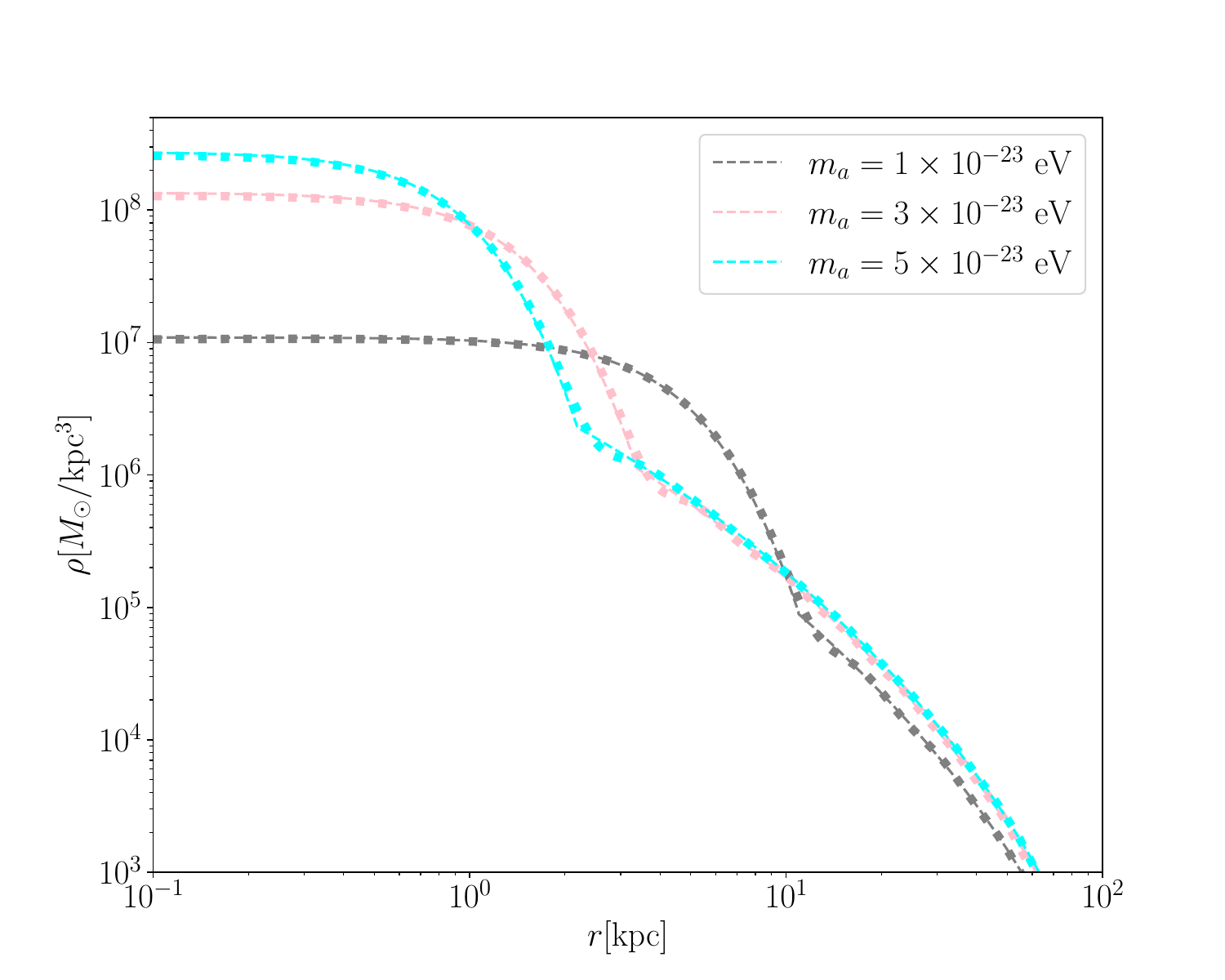}
  \caption{The radial profiles of the three FDM halos under consideration. The dashed lines represent the target FDM profiles $\rho_\text{in}(r)$ expressed in Eq. \ref{target}, which serve as the input for generating the initial wave functions. The parameters $k,~r_s,~r_\text{cut}$ and the halo mass within $r_\text{cut}$ are set to be $3,~10~$kpc, $50~$kpc and $10^{10}M_\odot$, respectively. The squares represent the reproduced density profiles. The halos corresponding to FDM mass values of $m_a/10^{-23}$eV values equal to $1,3,$ and $5$ are represented by gray, pink, and cyan colors, respectively.}
  \label{figure_0}
\end{figure}

We proceed to construct the halos using the methodology introduced by Yavetz et al. in \cite{Yavetz:2021pbc}. Initially, we solve the time-independent Schr$\ddot{\text{o}}$dinger equation under the static gravitational potential $\Phi_\text{in}(r)$, which is determined by the input target profile $\rho_\text{in}(r)$. We obtain the eigenstates  $\Psi_{nlm}(\mathbf{x})$ along with their corresponding eigenvalues $E_{nl}$. The expression for $\Psi_{nlm}(\mathbf{x})$ involves the radial wave function and spherical harmonic function 
\begin{equation}
    \Psi_{nlm}(\mathbf{x})=R_{nl}(r)Y^{m}_{l}(\theta,\varphi),
\end{equation}
where $n,l$, and $m$ denote the number of nodes in $R_{nl}$, angular, and magnetic quantum numbers, respectively. The time-dependent FDM wave function can be approximately written as a linear combination of these eigenstates \cite{Yavetz:2021pbc}
\begin{equation}
\psi_A(t,\mathbf{x})=\sum_{nlm}|a_{nl}|e^{i\phi_{nlm}}\Psi_{nlm}(\mathbf{x})e^{-iE_{nl}t/\hbar},
\label{psi_A}
\end{equation}
where $\phi_{nlm}$ denotes random initial phases within the range $[0,2\pi)$. The coefficients $|a_{nl}|$ are determined to ensure that the random phase averaged radial profile $\rho_\text{out}(r)=\frac{m_a}{4\pi}\sum_{nl}(2l+1)|a_{nl}|^2R^2_{nl}(r)$ aligns with the target input profile $\rho_\text{in}(r)$. The detailed method of obtaining $\Psi_{nlm}(\mathbf{x})$,  $|a_{nl}|$, and $E_{nl}$ can be found in Refs.~\cite{Yavetz:2021pbc,Yang}. The profiles reproduced from the constructed initial wave functions for the three halos are illustrated in Fig. \ref{figure_0} using squares. The halos corresponding to FDM mass values of $m_a/10^{-23}$eV values equal to $1,3,$ and $5$ are represented by grey, pink, and cyan colors, respectively. The dashed lines represent the target FDM profiles $\rho_\text{in}(r)$ as defined in Eq. \ref{target}, which served as input for generating the initial wave functions. It is evident that the output profiles aligns well with the target profiles.

It is crucial to highlight that the time-dependent wave function described by Eq. \ref{psi_A} is derived assuming a static isotropic potential. In real FDM halos, the presence of soliton oscillation, random walk, and dynamic granular fluctuations means that Eq. \ref{psi_A} is only an approximation. The precision of this approximation diminishes, especially as the halo evolves over extended periods \cite{Yavetz:2021pbc}. Consequently, we solely employ the wave function at $t=0$ as the initial condition for our simulation, 
expressed as
\begin{equation}
\psi(0,\mathbf{x})=\sum_{nlm}|a_{nl}|e^{i\phi_{nlm}}\Psi_{nlm}(\mathbf{x}).
\label{psi}
\end{equation}



\begin{figure*}
  \includegraphics[width=0.32\textwidth]{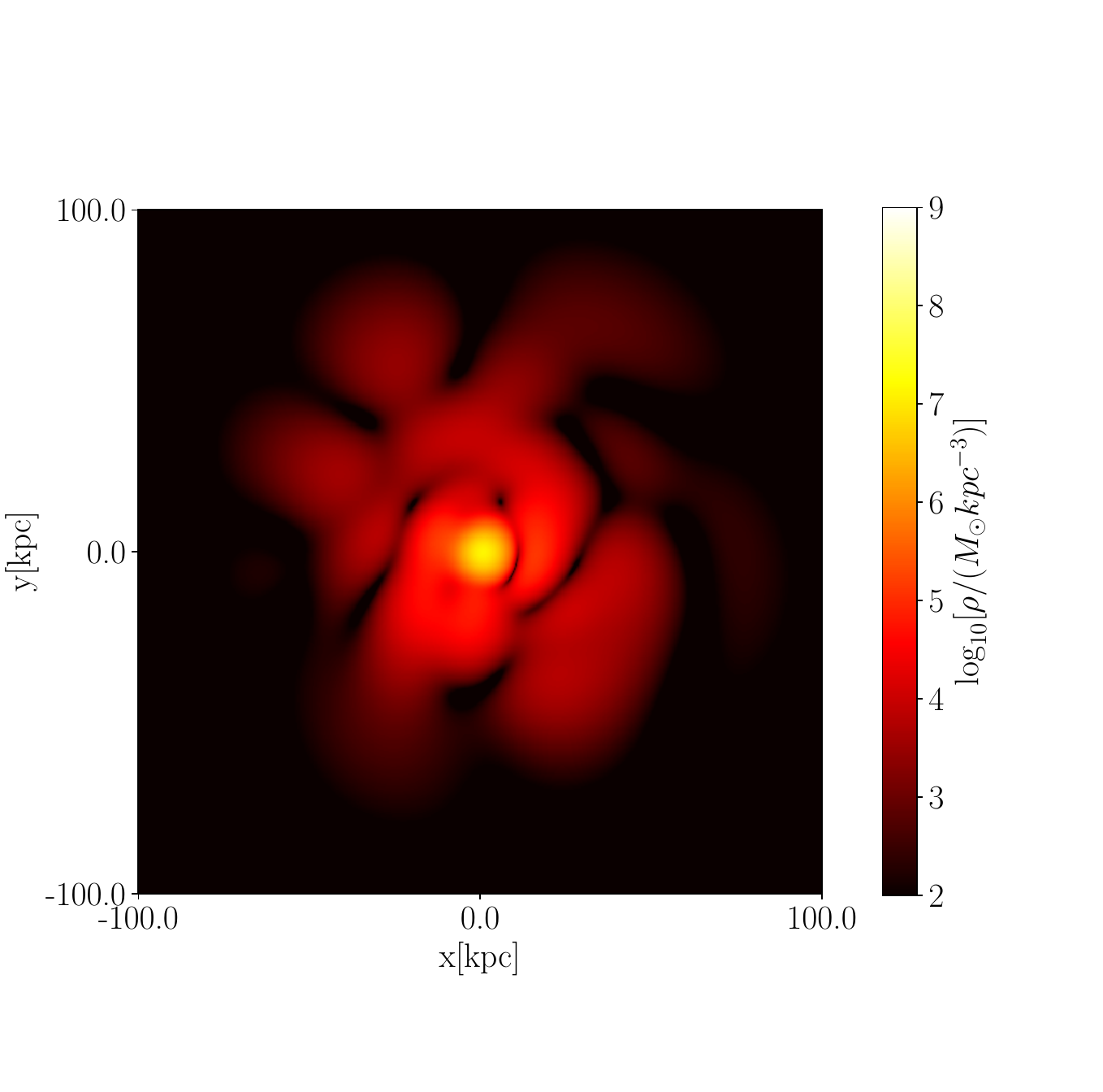}
  \includegraphics[width=0.32\textwidth]{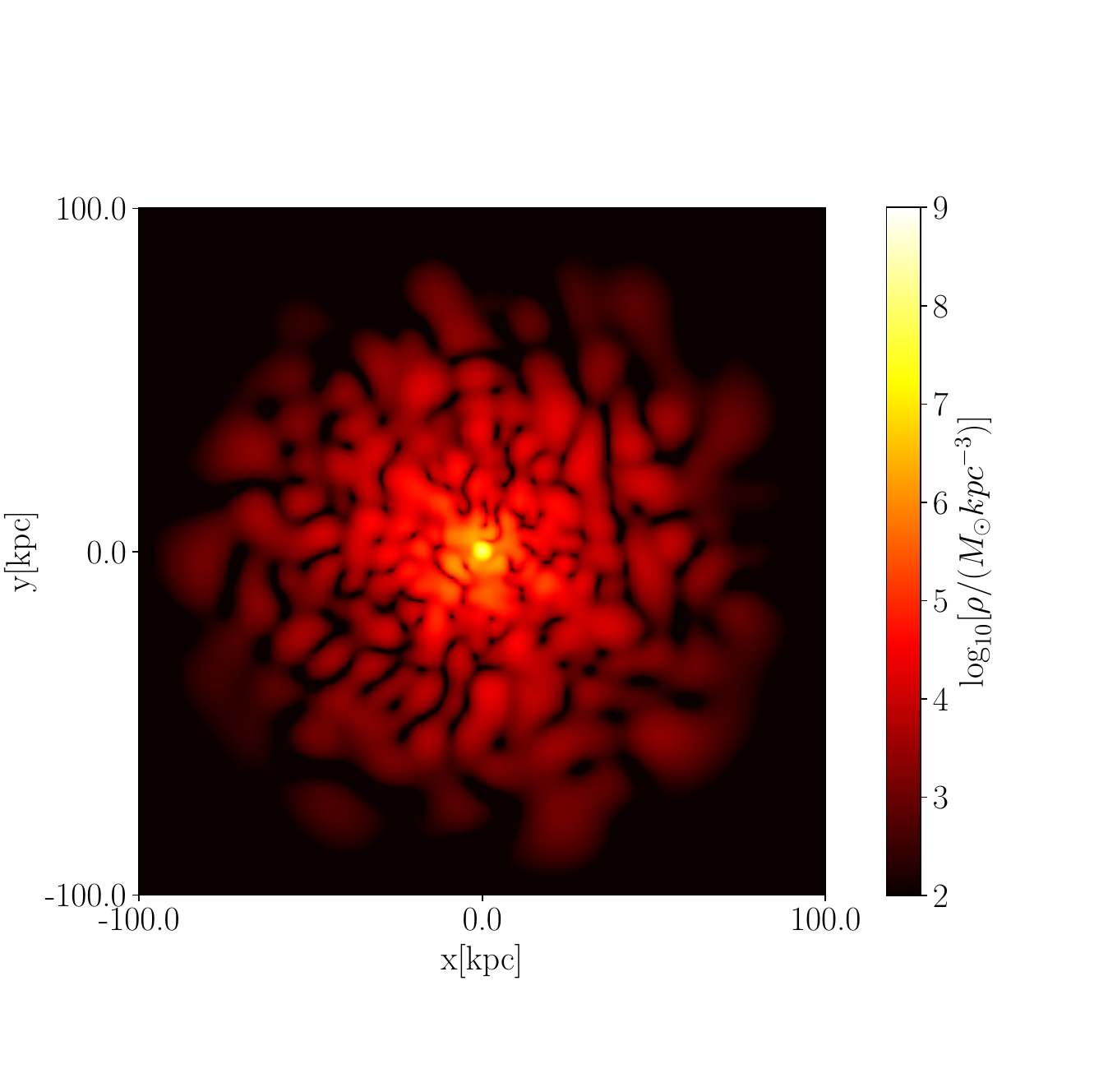}
  \includegraphics[width=0.32\textwidth]{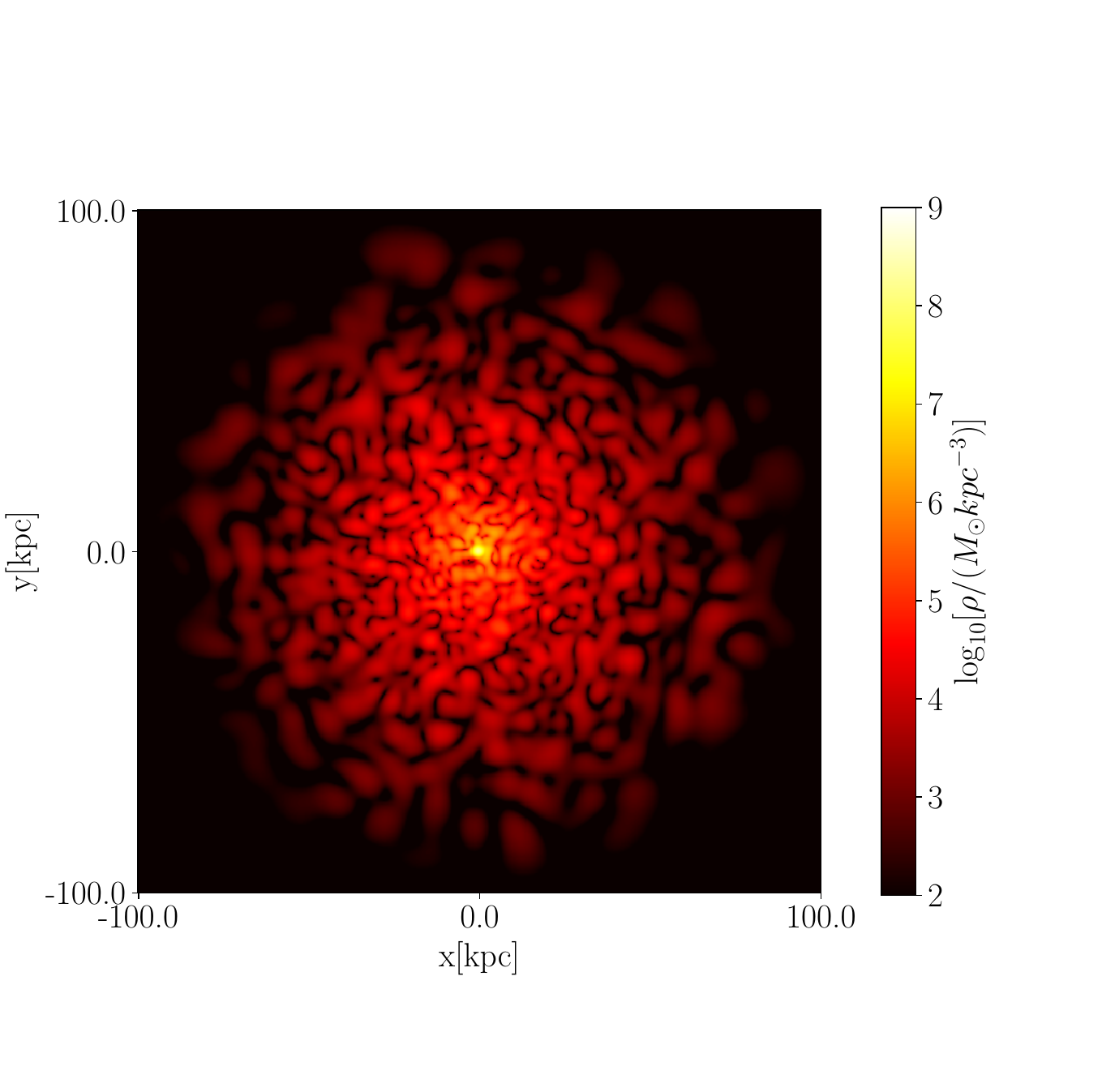}
  \caption{The initial density field $\rho(0,\mathbf{x})=m_a|\psi(0,\mathbf{x})|^2$ in the $z=0$ plane for the three constructed halos is displayed in the three panels, arranged in order of increasing $m_a/10^{-23}\text{eV}$ from left to right, corresponding to values of 1, 3, and 5, respectively. The form of the target profile for generating the initial wave functions for these halos is given by Eq. \ref{target}, with the parameters $k,~r_s,~r_\text{cut}$ and the halo mass within $r_\text{cut}$ set to be $3,~10~$kpc, $50~$kpc and $10^{10}M_\odot$, respectively. For all of these three halos, we only consider the eigenstates with eigenenergies below the energy of a particle on a circular orbit at $r_\text{cut}=50$ kpc.}
  \label{figure_1}
\end{figure*}

In our research, we only consider the eigenstates with eigenenergies below a maximum energy cutoff, which is assumed to be the energy of a particle on a circular orbit at $r_\text{cut}=50$ kpc. The permissible number of eigenstates are $136,~3554$, and $16125$ corresponding to $m_a/10^{-23}\text{ eV}=1,~3$, and $5$, respectively. This decreasing trend in the number of eigenstates indicates a reduction in quantum effect as the particle mass increases. The initial density distributions $\rho(0,\mathbf{x})=m_a|\psi(0,\mathbf{x})|^2$ in the $z=0$ plane for the constructed halos are illustrated in Fig. \ref{figure_1}. These visual results demonstrate that as the particle mass increases, there is a simultaneous reduction in both the size of the soliton core and the granularity in the outer NFW region, accompanied by an elevation in the core density.

While the choice of the energy cutoff does not affect the calculation of initial velocity in our analysis, it could have some impacts on the FDM density field, particularly in the outer region of the halo. Note that the energy cutoff we select is the energy of a particle on a circular orbit at $r_\text{cut} = 50$ kpc, while the virial radii of the three halos under consideration are approximately 43.9, 42.8, and 42.4 kpc, respectively, all of which are close to $r_\text{cut}$. Therefore, the results obtained through our approach would not differ significantly from those obtained by considering all eigenstates with eigenenergies below the gravitational potential energy at the virial radius \cite{Lin:2018whl}.

We employ the derived $\psi(0,\mathbf{x})$ as initial conditions, and evolve isolated FDM halos in simulations.\footnote{In cosmological simulations, the core of the halo would grow over time as the halo continuously attracts surrounding FDM \cite{Chen:2020cef} In this study, we do not consider this effect and treat the halo as an isolated system.} During the evolution, the FDM halos exhibit various effects that are absent in the CDM scenario, such as solition oscillation and random walk effects. We utilize the package  \textsc{PyUltraLight} \cite{Edwards:2018ccc}, which adopts the pseudo-spectral method, to evolve the FDM wave function, and enforce periodic boundary conditions within the simulation box. To mitigate boundary effects, we set the half length of one side of the simulation box to be $100$ kpc, which is significantly larger than the virial radius of the halos. The time step and resolution of our simulations are chosen to be $0.909$ Myr and $512^3$, respectively. It has been verified that our simulation results remain robust even with a smaller time step or higher resolution. Throughout our simulations of all three halos, we observe directional shifts of the center of mass, indicating an initial global velocity of the system described by $\psi(0,\mathbf{x})$. The movements of the center of mass in the $x$, $y$, and $z$ directions are, respectively, depicted by the blue, orange, and green dots in Fig. \ref{figure_2}.

\section{Theoretical interpretation of the initial global velocity \label{Sec_3}}
The initial global velocity associated with the FDM wave function $\psi(0,\mathbf{x})$ can be obtained by calculating its global momentum, which can be expressed in a form similar to that found in quantum mechanics
\begin{equation}
    \mathbf{P}=\int \psi^\star(-i\hbar\nabla)\psi d^3\mathbf{x}=-\frac{i\hbar}{2}\int(\psi^\star\nabla\psi-\psi\nabla\psi^\star)d^3\mathbf{x}.
    \label{P}
\end{equation}
However, it is important to clarify that, in fact, $\psi$ in this context represents a classical field, rather than a quantum mechanical wave function. The expression in Eq. \ref{P} can be comprehended through the fluid description of FDM \cite{hui2021wave}. 


\subsection{An intuitive understanding\label{understanding}}

Before rigorously calculating the nonzero initial global velocity of the constructed halos described by $\psi(0,\mathbf{x})$, we first aim to provide an intuitive understanding of its emergence. By showcasing that the system described by the wave function in Eq. \ref{psi_A} possesses a nonzero initial global velocity, we can understand the source of the nonzero global velocity carried by 
the initial wave function in Eq. \ref{psi}, which is also the velocity observed in the simulations.

In Sec. \ref{Sec_2}, we emphasized that Eq. \ref{psi_A} is an approximate wave function of the constructed halo, with its accuracy limited to a short time evolution duration. Looking from another perspective, Eq. \ref{psi_A} represents the exact solution of the Schr$\ddot{\text{o}}$dinger equation under the static potential $\Phi_\text{in}(r)$, which is assumed to be independent of $\psi_A(t,\mathbf{x})$ and remains constant with time. Therefore, the evolution described by Eq. \ref{psi_A} illustrates the evolution of the wave function under the static external potential $\Phi_\text{in}(r)$, rather than the evolution of an isolated halo wave function. We elucidate that the center of mass of the system described by Eq. \ref{psi_A} undergoes oscillations over time, indicating the existence of a nonzero initial global velocity at $t=0$.

This oscillatory motion of the center of mass arises by the interference between states of odd and even parity. This occurrence can be intuitively understood by considering the properties of spherical harmonic functions $Y_l^m(\pi-\theta,\pi+\varphi)=(-1)^lY_l^m(\theta,\varphi)$. 
When an odd parity state and an even parity state exhibit constructive interference at a position $\bm{r}$ at a particular moment, they must undergo destructive interference at $-\bm{r}$. Since the frequencies of these two states are different, the locations of interference enhancement and cancellation may interchange as time progresses, leading to the oscillatory motion of the center of mass.

This occurrence can also be rigorously elucidated by calculating the position of the center of mass $\mathbf{x}_c$ as
\begin{equation}
    \mathbf{x}_c=\int \mathbf{x}\, m_a\psi_A^\star\psi_A d^3\mathbf{x}/M_\text{halo},
\end{equation}
where $M_\text{halo}$ represents the total mass of the FDM halo
\begin{equation}
    M_\text{halo}=\int m_a\psi_A^\star\psi_A d^3\mathbf{x}=m_a\sum_{nlm}|a_{nlm}|^2.
\end{equation}
{As an example, we focus on $\mathbf{x}_c$ at the $z$ coordinate. The expression for $z_c$, as detailed in Appendix. \ref{A}, is provided as}
\begin{equation}
\begin{aligned}
z_c=\frac{m_a}{2M_\text{halo}}\sum_{n_1,l_1;n_2,l_2}&\frac{2}{\sqrt{(2l_1+1)(2l_2+1)}}|a_{n_1l_1}||a_{n_2l_2}|\\
&\times\int_0^\infty r^3R_{n_1l_1}R_{n_2l_2}\ dr\ \mathbf{\text{\large\MakeUppercase{\romannumeral 1}}}_z,
\end{aligned}
\label{z_c}
\end{equation}
where the factor of $1/2$ is included to avoid the repeated summation, and the expression of $\mathbf{\text{\large\MakeUppercase{\romannumeral 1}}}_z$ is detailed in Appendix. \ref{A}. The expression of $\mathbf{\text{\large\MakeUppercase{\romannumeral 1}}}_z$, which is proportional to $\delta_{l_1, l_2-1}$ or $\delta_{l_1-1, l_2}$, clearly reveals that only the interference between states with angular quantum numbers differing by one contributes to $z_c$, aligning with our intuitive understanding to some extent. States with angular quantum numbers differing by other odd numbers do not contribute to $z_c$. Nonetheless, the interference between these states contributes to $x_c$ and $y_c$, as can be observed from the momentum expression derived in the subsequent section.

\begin{figure*}
  \includegraphics[width=0.32\textwidth]{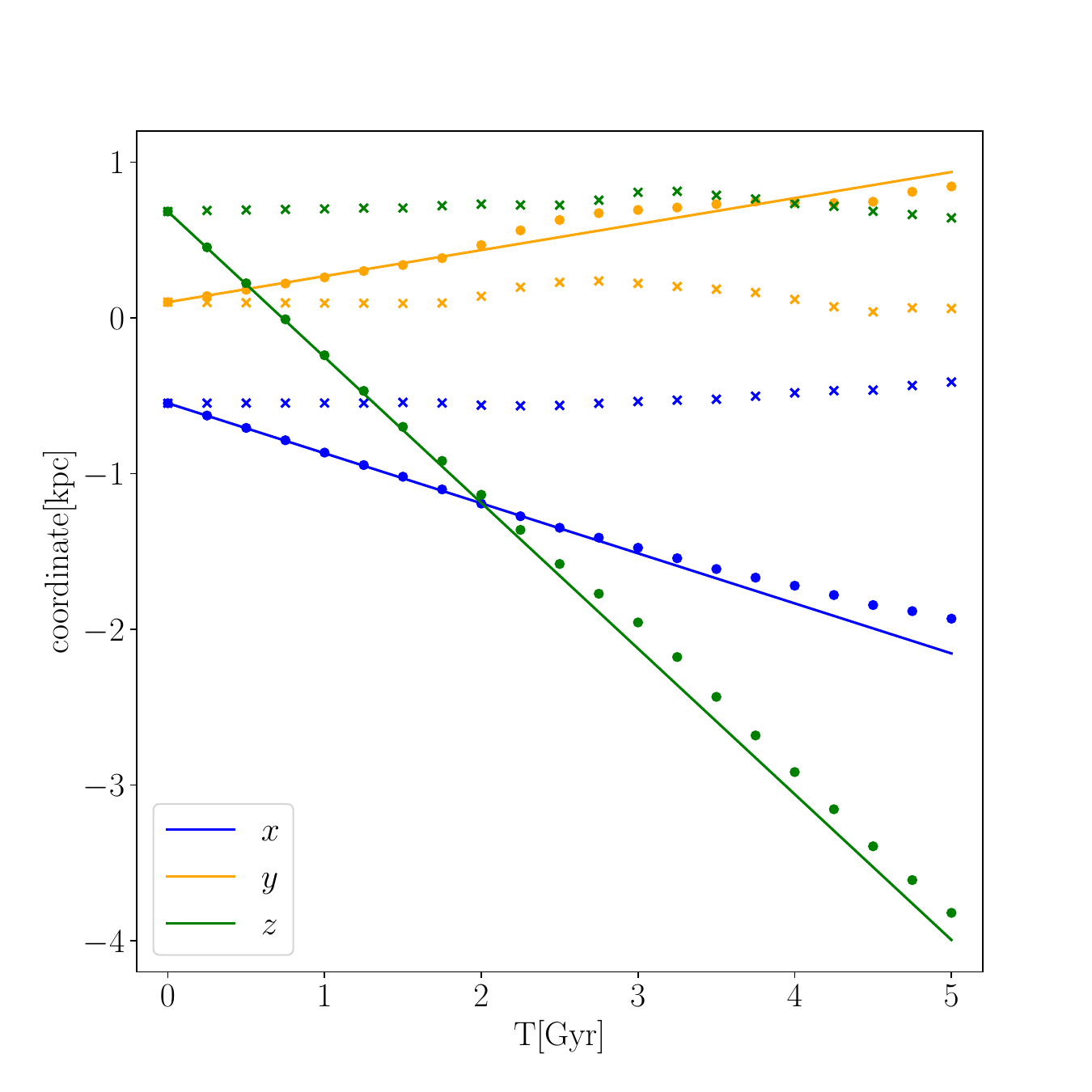}
  \includegraphics[width=0.32\textwidth]{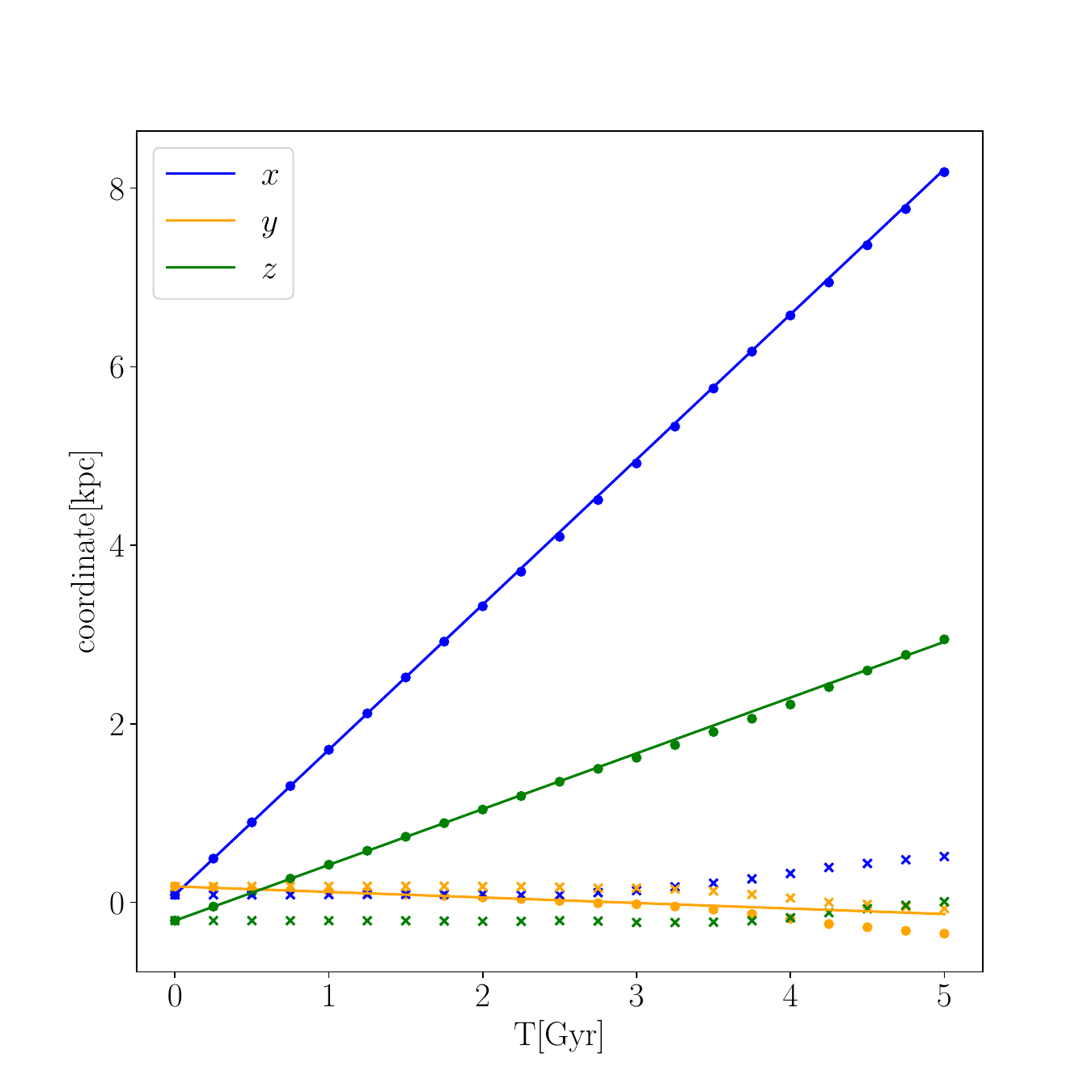}
  \includegraphics[width=0.32\textwidth]{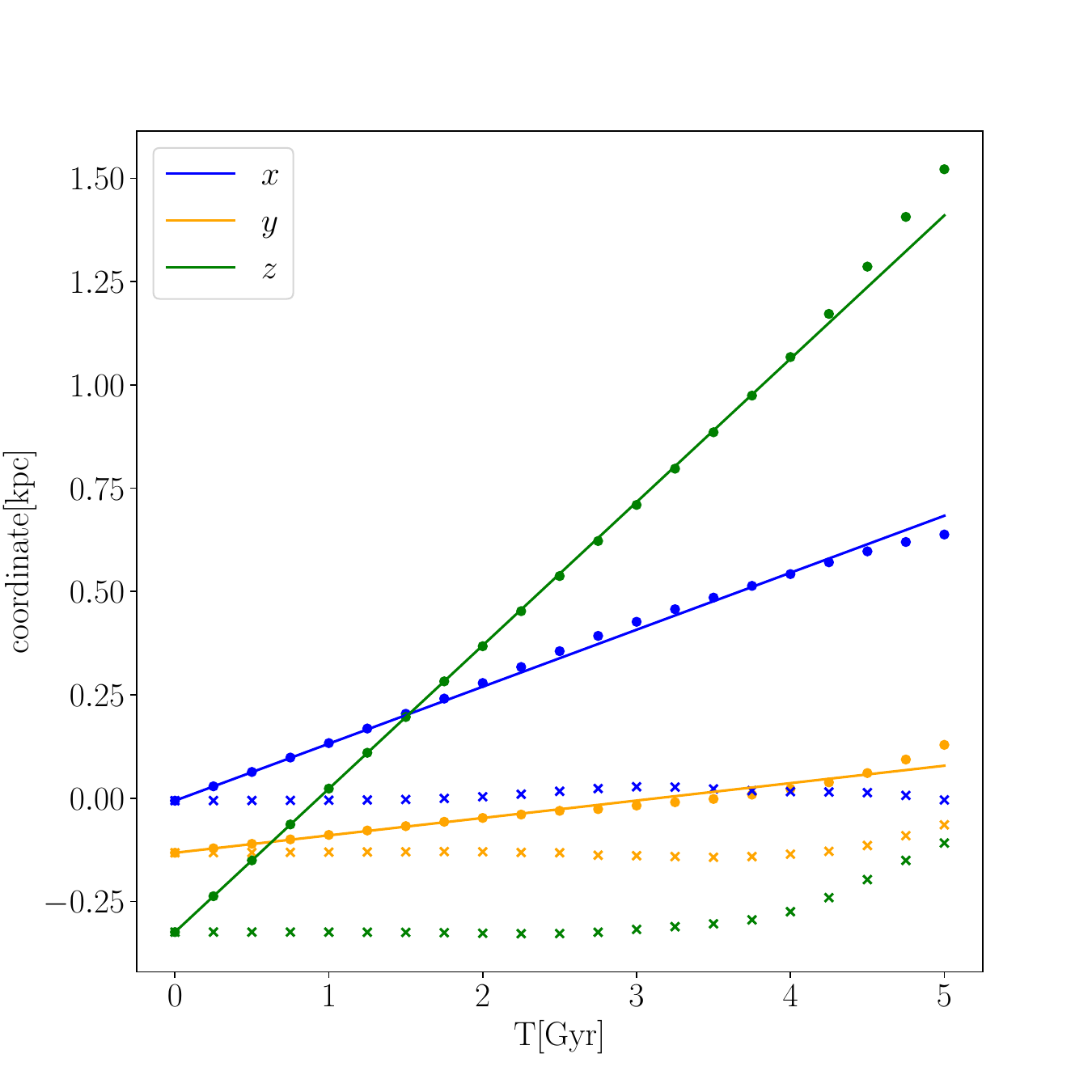}
  \caption{The center of mass motion of the three constructed halos are shown in the three panels, corresponding to $m_a/10^{-23}\text{ eV}=1,3,5$ from left to right, respectively. The blue, orange, and green colors represent the $x,y$, and $z$ coordinates, respectively. The solid lines and dots represent the theoretical results calculated according to Eq. \ref{v} and the simulation results for the original wave functions without boosting. The crosses represent the simulation results for the wave functions after boosting.}
  \label{figure_2}
\end{figure*}

\subsection{Formula of the initial velocity}
In this subsection, we present the results of the calculation of the global velocity carried by our initial wave function in Eq. \ref{psi}. Following a series of calculations detailed in Appendix \ref{B}, we derive the three components of the system's initial momentum, given in Eq. \ref{P}, as follows:
\begin{equation}
\begin{aligned}
    P_i=&\frac{1}{2}\sum_{n_1,l_1;n_2,l_2}\frac{-\hbar}{\sqrt{(2l_1+1)(2l_2+1)}}|a_{n_1l_1}||a_{n_2l_2}|\\
    &\times\left[\int_0^\infty \left(R_{n_1l_1}\partial_rR_{n_2l_2}-R_{n_2l_2}\partial_rR_{n_1l_1}\right)r^2dr\ \mathbf{\text{\large\MakeUppercase{\romannumeral 1}}}_{P_i}\right.\\
    &\hspace{3cm}\left.+\int_0^\infty rR_{n_1l_1}R_{n_2l_2}dr\ \mathbf{\text{\large\MakeUppercase{\romannumeral 2}}}_{P_i}\right],
\end{aligned}
\label{P_i}
\end{equation}
where $i=x,~y,~z$ and the expressions of $\mathbf{\text{\large\MakeUppercase{\romannumeral 1}}}_{P_i}$ and $\mathbf{\text{\large\MakeUppercase{\romannumeral 2}}}_{P_i}$ are detailed in Appendix. \ref{B}. While $P_z$ solely receives contributions from the interference of states with $l$ differing by one, the expressions of $P_x$ and $P_y$ demonstrate contributions not only from the interference of states with $l$ differing by one but also from the interference of states with $l$ differing by other odd numbers. In the expressions of $\mathbf{\text{\large\MakeUppercase{\romannumeral 2}}}_{P_x}$ and $\mathbf{\text{\large\MakeUppercase{\romannumeral 2}}}_{P_y}$, there are terms proportional to $W(k_1,k_2,m)$, which is expressed as
\begin{equation}
\begin{aligned}
    W(k_1,k_2,m)&=\sum_k\left[1+(-1)^k\right]\sqrt{\frac{(k-2)!}{(k+2)!}}(2k+1)\\
    &\quad\times\begin{pmatrix}
k_1 & k_2 & k \\
0 & 0 & 0 \\
\end{pmatrix}\begin{pmatrix}
k_1 & k_2 & k \\
-m & m+2 & -2 \\
\end{pmatrix},
\end{aligned}
\end{equation}
where $|k_1-k_2|=|l_1-l_2-1|$ and $\begin{pmatrix}
\cdot & \cdot & \cdot \\
\cdot & \cdot & \cdot \\
\end{pmatrix}$ represents the Wigner 3-j symbol \cite{edmonds1996angular}. The factor $1+(-1)^k$ indicates that only the terms with even $k$ contribute to $W$. Considering the properties of the Wigner 3-j symbol  $\begin{pmatrix}
k_1 & k_2 & k \\
0 & 0 & 0 \\
\end{pmatrix}$ \cite{edmonds1996angular}, $W$ is nonzero only when $k_1+k_2+k$ is an even number, which requires that $k_1+k_2$ is an even number. This implies that only cases where $l_1$ and $l_2$ differ by an odd number contribute to $P_x$ and $P_y$. This is consistent with the intuitive understanding presented in Sec. \ref{understanding}.

The initial global velocity of the constructed FDM halo can be obtained by dividing $\mathbf{P}$ by the total halo mass
\begin{equation}
    \mathbf{v}=\frac{\mathbf{P}}{M_\text{halo}}.
    \label{v}
\end{equation}
The predicted motion of the center of mass in the $x,~y$, and $z$ directions calculated using Eqs.~\ref{P_i} and \ref{v} is illustrated in Fig. \ref{figure_2} by the solid blue, orange, and green lines, respectively. The alignment between our calculated predictions and the simulation results is visually evident from these illustrations.


\begin{figure*}[htbp]
  \includegraphics[width=0.99\textwidth]{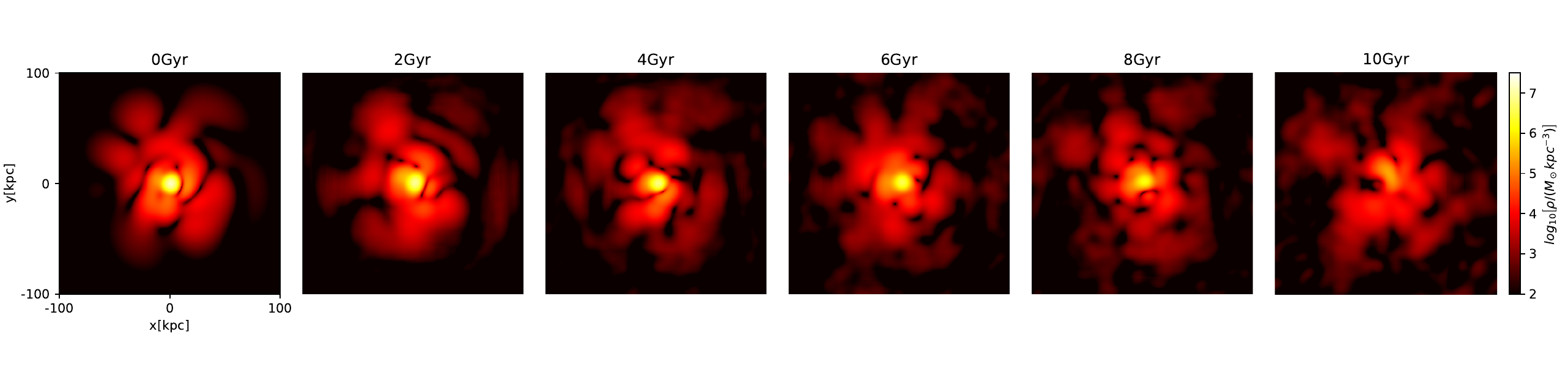}
  \includegraphics[width=0.99\textwidth]{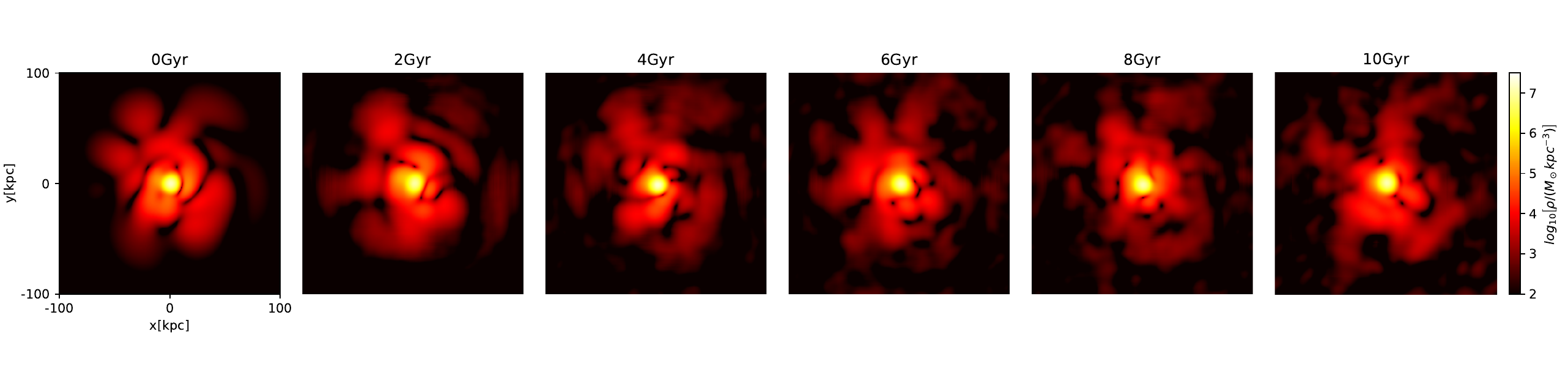}
  \caption{Evolution of the 3D density field on the $z=0$ plane without (top row) and with (bottom row) boosting, using $m_a/10^{-23}\text{ eV}=1$ as an example.}
  \label{figure_3}
\end{figure*}

\section{Galilean boost\label{Sec_4}}
The emergence of a nonzero initial global velocity may seem strange at first glance, as our construction process begins with a spherically symmetric and isotropic density profile, which should not imply a velocity directed in any specific direction. However, the issue arises from the need to construct a complex wave function from a real density profile, thereby introducing additional degrees of freedom in the process. This can be understood through the following argument: the density field can only determine the distribution of the wave function's magnitude but cannot dictate the distribution of the wave function's phase. 

If we simply take the wave function as $\sqrt{\rho_\text{in}/m}e^{i\alpha}$ and substitute it into Eq. \ref{P}, the result would indeed be zero. However, in the construction process we discussed, we express the wave function as a summation of eigenstates, each assigned a random phase. This operation causes the wave function with more information than the initial input density distribution, consequently resulting in a nonzero initial velocity. From a different perspective, if we require the constructed wave function in the initial conditions to not only produce the desired density profile but also satisfy a constraint of zero global velocity, then the phases of these eigenstates cannot be randomly chosen. Instead, they should be adjusted to ensure that the calculated initial velocity represented by Eq. \ref{v} is zero. However, due to the involvement of a large number of states, this adjustment becomes challenging in practical implementation.

In this study, we utilize an effective method involving a boost to eliminate the undesired initial velocity. Specifically, we can obtain a new wave function $\tilde{\psi}_A(t,\mathbf{x})$ in a uniformly moving reference frame with a velocity equal to the initial global velocity of $\psi_A(t,\mathbf{x})$. Then, by using $\tilde{\psi}_A(0,\mathbf{x})$ as the initial condition for a new simulation, the initial global velocity is effectively removed. Assuming the initial global velocity of a constructed halo is $\mathbf{v}$, a simple substitution $\psi_A(t,\mathbf{x})\to \psi_A(t,\mathbf{x}+\mathbf{v}t)$ is not sufficient, as this new wave function no longer satisfies the SP equations under the static input potential. In Appendix \ref{C}, we demonstrate that we can ensure the new wave function still obeys the SP equations by making the following substitution \cite{Edwards:2018ccc}
\begin{equation}
    \psi_A(t,\mathbf{x})\to\widetilde{\psi}_A(t,\mathbf{x})\equiv\psi_A(t,\mathbf{x}+\mathbf{v}t)e^{i(-m_a\mathbf{v}\cdot\mathbf{x}-\frac{1}{2}m_a\mathbf{v}^2t)/\hbar}.
\end{equation}

Subsequently, we utilize the new wave function at $t=0$ as an initial condition and carry out a new series of simulations. Our results show that the center of mass remains relatively stationary within the resolution range, as illustrated in Fig. \ref{figure_2} by the crosses. Figure \ref{figure_3} displays the evolution of the 3D density field on the $z=0$ plane without (top row) and with (bottom row) boosting, with $m_a/10^{-23}\text{ eV}=1$ as an example. In the scenario without boosting, the $z$ component of the halo mass center velocity is a significantly large negative value, as shown in the left panel of Fig. \ref{figure_2}, and the soliton undergoes a random walk around the center of mass. As time progresses, the soliton's outline becomes less distinct and eventually disappears at $10$ Gyr due to its departure from the $z=0$ plane. Conversely, when the boost is applied, the center of mass remains stationary while the soliton continues its random walk around the center, maintaining a clear outline throughout the evolution.

By following a similar approach, we can further manipulate the new wave function $\widetilde{\psi}_A$ to construct FDM halos with arbitrary initial velocities.

\section{Conclusion\label{Sec_5}}

In this study, we have elucidated how the interference between states of odd and even parity in the FDM wave function can result in an initial global velocity of a constructed halo. We have presented the formula for the initial global velocity, which aligns well with the simulation results. In principle, this initial global velocity could be eliminated by adjusting phases in the initially constructed FDM wave function. However, practical implementation of this approach is challenging. Instead, we have employed an effective method to counteract this velocity through a Galilean boost, enabling the creation of a stationary FDM halo using the theoretical expression. This approach can also be extended to generate halos with arbitrary initial global velocities, facilitating controlled FDM simulations for studying tidal effects, galaxy collisions, and other phenomena in galaxies.

\acknowledgments
This work is supported by the National Natural Science Foundation of China under Grants No. 12447105 and 12175248.


\bibliography{FDM}
\begin{appendix}
\begin{widetext}
\section{Derivation of the mass center oscillation\label{A}}
In this appendix, we illustrate the calculation of the center of mass location of the constructed FDM halo using the $z$ coordinate as an example. Initially, we focus on two states in the sum of Eq. \ref{psi_A}
\begin{equation}
\psi_A^{(12)}(t,\mathbf{x})=|a_1|e^{i\phi_1}R_1(r)Y_1(\theta,\varphi)e^{-iE_1t/\hbar}+|a_2|e^{i\phi_2}R_2(r)Y_2(\theta,\varphi)e^{-iE_2t/\hbar},
\end{equation}
where we use 1 and 2 to differentiate between these two states. The contribution of these two states to $z_c$ is given by
\begin{equation}
\begin{aligned}
    M_\text{halo}z^{(12)}_c/m_a&=\int z\psi_A^{(12)\star}(t,\mathbf{x})\psi_A^{(12)}(t,\mathbf{x})d^3\mathbf{x}\\
    &=\int r^3\sin\theta\cos\theta\left[|a_1|^2R^2_1|Y_1|^2+|a_2|^2R^2_2|Y_2|^2\right.\\
    &\left.\quad+|a_1||a_2|e^{-i(\phi_1-\phi_2)+i(E_1-E_2)t/\hbar}R_1R_2Y_1^{\star}Y_2+|a_1||a_2|e^{i(\phi_1-\phi_2)-i(E_1-E_2)t/\hbar}R_1R_2Y_1Y_2^{\star}\right]drd\theta d\varphi.
\end{aligned}
\label{Az_c}
\end{equation}
The first term can be expressed as
\begin{equation}
    \begin{aligned}
        \int_0^\infty r^3\sin\theta\cos\theta\ |a_1|^2R^2_1|Y_1|^2\propto \int_0^\pi P^{m_1}_{l_1}(\cos\theta)P^{m_1}_{l_1}(\cos\theta)\sin\theta\cos\theta d\theta,
    \end{aligned}
\end{equation}
which can be shown to be zero using the following two identities of the associated Legendre polynomials \cite{2015867}
\begin{equation}
        \int_{-1}^1P^m_l(x)P^m_k(x)dx=\frac{(l+m)!}{(l-m)!}\frac{2}{2l+1}\delta_{l,k},
        \label{P_1}
\end{equation}
\begin{equation}
    (2l+1)xP^m_l(x)=(l+1)P^m_{l-1}(x)+(l-m+1)P^m_{l+1}(x),
    \label{P_2}
\end{equation}
where $\delta_{l,k}$ represents the Kronecker $\delta$ function. Similarly, the second term of Eq. \ref{Az_c} is also zero. The two terms in the second line of Eq. \ref{Az_c} can be expressed as follows
\begin{equation}
    \begin{aligned}
        &|a_1||a_2|(-1)^{m_1}\sqrt{\frac{2l_1+1}{4\pi}\frac{(l_1-m_1)!}{(l_1+m_1)!}}(-1)^{m_2}\sqrt{\frac{2l_2+1}{4\pi}\frac{(l_2-m_2)!}{(l_2+m_2)!}}\int_0^{2\pi} 2\cos\left[\phi_1-\phi_2-(E_1-E_2)t/\hbar+(m_1-m_2)\varphi\right]d\varphi\\
        &\hspace{8cm}\times\int_0^\infty r^3R_1R_2 dr\int_0^\pi P^{m_1}_{l_1}(\cos\theta)P^{m_2}_{l_2}(\cos\theta)\sin\theta\cos\theta d\theta\\
        =&|a_1||a_2|(-1)^{m_1}\sqrt{\frac{2l_1+1}{4\pi}\frac{(l_1-m_1)!}{(l_1+m_1)!}}(-1)^{m_2}\sqrt{\frac{2l_2+1}{4\pi}\frac{(l_2-m_2)!}{(l_2+m_2)!}}4\pi\delta_{m_1.m_2}\cos\left[\phi_1-\phi_2-(E_1-E_2)t/\hbar\right]\\
        &\hspace{8cm}\times\int_0^\infty r^3R_1R_2 dr\int_0^\pi P^{m_1}_{l_1}(\cos\theta)P^{m_2}_{l_2}(\cos\theta)\sin\theta\cos\theta d\theta.\\
    \end{aligned}
\end{equation}
By utilizing Eqs. \ref{P_1} and \ref{P_2} again, we can simplify Eq. \ref{Az_c} to the form 
\begin{equation}
    \begin{aligned}
        M_\text{halo}z^{(12)}_c/m_a&=\frac{2}{\sqrt{(2l_1+1)(2l_2+1)}}\delta_{m_1,m_2}|a_1||a_2|\cos\left[\phi_1-\phi_2-(E_1-E_2)t/\hbar\right]\int_0^\infty r^3R_1R_2dr\\
        &\hspace{3cm}\times\left[\sqrt{(l_2-m_2)(l_2+m_2)}\delta_{l_1,l_2-1}+\sqrt{(l_1-m_1)(l_1+m_1)}\delta_{l_1-1,l_2}\right].
    \end{aligned}
\end{equation}
Summing over all states' contributions, we can obtain Eq. \ref{z_c}. The expression of $\mathbf{\text{\large\MakeUppercase{\romannumeral 1}}}_z$ is given by
\begin{equation}
    \begin{aligned}
    \mathbf{\text{\large\MakeUppercase{\romannumeral 1}}}_z=\sum_{m=\text{max}\{-l_1,-l_2\}}^{\text{min}\{l_1,l_2\}}&\cos\left[\phi_{n_1l_1m}-\phi_{n_2l_2m}-\left(E_{n_1l_1}-E_{n_2l_2}\right)t/\hbar\right]\\
    &\times\left[\sqrt{(l_2-m)(l_2+m)}\ \delta_{l_1,l_2-1}+\sqrt{(l_1-m)(l_1+m)}\ \delta_{l_1-1,l_2}\right].
    \end{aligned}
\end{equation}

\section{Derivation of the initial momentum\label{B}}
In this appendix, we calculate the initial momentum of the system represented by the wave function of Eq. \ref{psi}. Similar to our previous approach, we first consider the contribution of two states and then sum over all states in Eq. \ref{psi}. By substituting $\psi^{(12)}(0,\mathbf{x})=|a_1|e^{i\phi_1}R_1(r)Y_1(\theta,\varphi)+|a_2|e^{i\phi_2}R_2(r)Y_2(\theta,\varphi)$ into Eq. \ref{P} and simplifying it through some rearrangements, we arrive at the following expression
\begin{equation}
    \begin{aligned}
        \mathbf{P}^{(12)}&=|a_1||a_2|(-\hbar)(-1)^{m_1}\sqrt{\frac{2l_1+1}{4\pi}\frac{(l_1-m_1)!}{(l_1+m_1)!}}(-1)^{m_2}\sqrt{\frac{2l_2+1}{4\pi}\frac{(l_2-m_2)!}{(l_2+m_2)!}}\\
        &\times\int\left\{\sin\left[\phi_1-\phi_2+(m_1-m_2)\varphi\right]\left[(R_1\partial_r R_2-R_2\partial_r R_1)P^{m_1}_{l_1}P^{m_2}_{l_2}\mathbf{e}_r+\frac{R_1R_2}{r}(P^{m_1}_{l_1}\partial_\theta P^{m_2}_{l_2}-P^{m_2}_{l_2}\partial_\theta P^{m_1}_{l_1})\mathbf{e}_\theta\right]\right.\\
        &\hspace{2cm}\left.-\cos\left[\phi_1-\phi_2+(m_1-m_2)\varphi\right]\frac{m_1+m_2}{r\sin\theta}R_1R_2P^{m_1}_{l_1}P^{m_2}_{l_2}\mathbf{e}_\varphi\right\}r^2\sin\theta drd\theta d\varphi.
    \end{aligned}
    \label{vec_P}
\end{equation}
We start our calculation by focusing on the third component of $\mathbf{P}$, which can be derived from the above equation as
\begin{equation}
    \begin{aligned}
        P^{(12)}_z&=|a_1||a_2|(-\hbar)\sqrt{\frac{2l_1+1}{4\pi}\frac{(l_1-m_1)!}{(l_1+m_1)!}}\sqrt{\frac{2l_2+1}{4\pi}\frac{(l_2-m_2)!}{(l_2+m_2)!}}2\pi\delta_{m_1,m_2}\sin(\phi_1-\phi_2)\\
        &\times\left\{\int_0^\infty(R_1\partial_rR_2-R_2\partial_rR_1)r^2dr\int_0^\pi P^{m_1}_{l_1}(\cos\theta)P^{m_2}_{l_2}(\cos\theta)\cos\theta\sin\theta d\theta\right.\\
        &\hspace{2cm}+\left.\int_0^\infty rR_1R_2 dr\int_0^\pi\left[P^{m_1}_{l_1}(\cos\theta)\partial_\theta P^{m_2}_{l_2}(\cos\theta)-P^{m_2}_{l_2}\partial_\theta P^{m_1}_{l_1}(\cos\theta)\right](-\sin^2\theta)d\theta \right\}.
    \end{aligned}
\end{equation}
Here, we have integrated out the integral of $\varphi$. Utilizing Eqs. \ref{P_1} and \ref{P_2} and the identity \cite{2015867}
\begin{equation}
    (2l+1)(1-x^2)\frac{dP^m_l(x)}{dx}=(l+1)(l+m)P^m_{l-1}(x)-l(l-m+1)P^m_{l+1}(x),
    \label{P_3}
\end{equation}
the integral with respect to $\theta$ can be evaluated. Subsequently, upon summing over all states' contributions, we obtain the expression of $P_z$ in the form of Eq. \ref{P_i}, and the expression of $\mathbf{\text{\large\MakeUppercase{\romannumeral 1}}}_{P_z}$ and $\mathbf{\text{\large\MakeUppercase{\romannumeral 2}}}_{P_z}$ are given by
\begin{equation}
    \mathbf{\text{\large\MakeUppercase{\romannumeral 1}}}_{P_z}=\sum_{m=\text{max}\left\{-l_1,-l_2\right\}}^{\text{min}\left\{l_1,l_2\right\}}\sin\left(\phi_{n_1l_1m}-\phi_{n_2l_2m}\right)\left[\sqrt{(l_2-m)(l_2+m)}\delta_{l_1,l_2-1}+\sqrt{(l_1-m)(l_1+m)}\delta_{l_1-1,l_2}\right],
\end{equation}
\begin{equation}
    \mathbf{\text{\large\MakeUppercase{\romannumeral 2}}}_{P_z}=\sum_{m=\text{max}\left\{-l_1,-l_2\right\}}^{\text{min}\left\{l_1,l_2\right\}}\sin\left(\phi_{n_1l_1m}-\phi_{n_2l_2m}\right)(l_1+l_2+1)\left[\sqrt{(l_2-m)(l_2+m)}\delta_{l_1,l_2-1}-\sqrt{(l_1-m)(l_1+m)}\delta_{l_1-1,l_2}\right].
\end{equation}
Similarly, we can derive the expression for $P_x$ from Eq. \ref{vec_P} as follows:
\begin{equation}
    \begin{aligned}
        P^{(12)}_x&=|a_1||a_2|(-\hbar)(-1)^{m_1}\sqrt{\frac{2l_1+1}{4\pi}\frac{(l_1-m_1)!}{(l_1+m_1)!}}(-1)^{m_2}\sqrt{\frac{2l_2+1}{4\pi}\frac{(l_2-m_2)!}{(l_2+m_2)!}}\pi\sin(\phi_1-\phi_2)\\
        &\times\left\{\int_0^\infty(R_1\partial_rR_2-R_2\partial_rR_1)r^2dr\int_0^\pi P^{m_1}_{l_1}(\cos\theta)P^{m_2}_{l_2}(\cos\theta)\sin^2\theta d\theta \left(\delta_{m_1,m_2-1}+\delta_{m_1-1,m_2}\right)\right.\\
        &\quad+\int_0^\infty rR_1R_2 dr\left[\int_0^\pi\left[P^{m_1}_{l_1}(\cos\theta)\partial_\theta P^{m_2}_{l_2}(\cos\theta)-P^{m_2}_{l_2}\partial_\theta P^{m_1}_{l_1}(\cos\theta)\right]\cos\theta\sin\theta d\theta\left(\delta_{m_1,m_2-1}+\delta_{m_1-1,m_2}\right) \right.\\
        &\hspace{5cm}\left.\left.+(m_1+m_2)\int_0^\pi P^{m_1}_{l_1}(\cos\theta)P^{m_2}_{l_2}(\cos\theta)d\theta\left(\delta_{m_1,m_2-1}-\delta_{m_1-1,m_2}\right)\right]\right\}.
    \end{aligned}
\end{equation}
We then integrate out $\theta$ using Eqs. \ref{P_1} and \ref{P_2} and the following identities for associated Legendre polynomials \cite{2015867,HarryAMavromatis_1999,DONG2002541}
\begin{equation}
    (2l+1)\sqrt{1-x^2}P^m_l(x)=P^{m+1}_{l-1}(x)-P^{m+1}_{l+1}(x),
\end{equation}
\begin{equation}
    \frac{1}{\sqrt{1-x^2}}P^m_l(x)=-\frac{1}{2m}\left[P^{m+1}_{l+1}(x)+(l-m+1)(l-m+2)P^{m-1}_{l+1}(x)\right]\quad (m\neq 0),
\end{equation}
\begin{equation}
    \sqrt{1-x^2}\frac{dP^m_l(x)}{dx}=\frac{1}{2}\left[(l+m)(l-m+1)P^{m-1}_l(x)-P^{m+1}_l\right],
\end{equation}
\begin{equation}
\begin{aligned}
    \int_{-1}^1P^m_{k_1}(x)P^{m+2}_{k_2}(x)dx&=(-1)^m 2\sqrt{\frac{(k_1+m)!(k_2+m+1)!}{(k_1-m)!(k_2-m-2)!}}\sum_k\left[1+(-1)^k\right]\sqrt{\frac{(k-2)!}{(k+2)!}}(2k+1)\\
    &\hspace{6cm}\times\begin{pmatrix}
k_1 & k_2 & k \\
0 & 0 & 0 \\
\end{pmatrix}\begin{pmatrix}
k_1 & k_2 & k \\
-m & m+2 & -2 \\
\end{pmatrix},
\end{aligned}
\end{equation}
where $\begin{pmatrix}
\cdot & \cdot & \cdot \\
\cdot & \cdot & \cdot \\
\end{pmatrix}$ represents the Wigner 3-j symbol \cite{edmonds1996angular}. Summing over all states, the expression of $P_x$ can be obtained in the form of Eq. \ref{P_i}, with
\begin{equation}
    \mathbf{\text{\large\MakeUppercase{\romannumeral 1}}}_{P_x}=\widetilde{\mathbf{\text{\large\MakeUppercase{\romannumeral 1}}}}_{P_x}(n_1,l_1;n_2,l_2)-\widetilde{\mathbf{\text{\large\MakeUppercase{\romannumeral 1}}}}_{P_x}(n_2,l_2;n_1,l_1),
\end{equation}
\begin{equation}
    \mathbf{\text{\large\MakeUppercase{\romannumeral 2}}}_{P_x}=\widetilde{\mathbf{\text{\large\MakeUppercase{\romannumeral 2}}}}_{P_x}(n_1,l_1;n_2,l_2)+\widetilde{\mathbf{\text{\large\MakeUppercase{\romannumeral 2}}}}_{P_x}(n_2,l_2;n_1,l_1),
\end{equation}
where 
\begin{equation}
\begin{aligned}
    \widetilde{\mathbf{\text{\large\MakeUppercase{\romannumeral 1}}}}_{P_x}(n_1,l_1;n_2,l_2)&=\sum_{m=\text{max}\left\{-l_1,-l_2-1\right\}}^{\text{min}\left\{l_1,l_2-1\right\}}\frac{1}{2}\sin[\phi_{n_1l_1m}-\phi_{n_2l_2(m+1)}]\\
    &\hspace{3cm}\times\left[\sqrt{(l_1+m+1)(l_2+m+1)}\delta_{l_1,l_2-1}-\sqrt{(l_1-m)(l_2-m)}\delta_{l_1-1,l_2}\right],
\end{aligned}
\end{equation}
\begin{equation}
    \begin{aligned}
     \widetilde{\mathbf{\text{\large\MakeUppercase{\romannumeral 2}}}}_{P_x}(n_1,l_1;n_2,l_2)&=\sum_{m=\text{max}\left\{-l_1,-l_2-1\right\}}^{\text{min}\left\{l_1,l_2-1\right\}}\frac{1}{4}\sin[\phi_{n_1l_1m}-\phi_{n_2l_2(m+1)}]\\
     &\times\left\{\sqrt{(l_1+m+1)(l_2+m+1)}\left[(l_1+l_2-2m)+\frac{2m+1}{m}(2l_1+1)(1-\delta_{m,0})\right]\delta_{l_1,l_2-1}\right.\\
     &\quad+\left[(l_1+l_2+2m+2)\sqrt{(l_1-m)(l_2-m)}+(2l_2+1)\sqrt{l_1l_2}\delta_{m,0}\right]\delta_{l_1-1,l_2}\\
     &\quad +(-1)^{m+1}(2l_2+1)\sqrt{(l_2-m-1)(l_2+m+2)}\left[\sqrt{(l_1-m)(l_1+m)}W(l_1-1,l_2,m)\right.\\
     &\qquad\left.+\sqrt{(l_1-m+1)(l_1+m+1)}W(l_1+1,l_2,m)\right]+(-1)^m(2l_1+1)\sqrt{(l_1-m+1)(l_1+m)}\\
     &\qquad\left[\sqrt{(l_2-m-1)(l_2+m+1)}W(l_1,l_2-1,m-1)+\sqrt{(l_2-m)(l_2+m+2)}W(l_1,l_2+1,m-1)\right]\\
     &\quad+(2l_1+1)(2l_2+1)\left[\sqrt{(l_2+2)(l_2+3)}\delta_{m,0}W(l_1,l_2+1,0)\right.\\
     &\qquad\left.\left.+(-1)^{m-1}\frac{2m+1}{m}\sqrt{(l_1-m+1)(l_1-m+2)}(1-\delta_{m,0})W(l_1+1,l_2,m-1)\right]\right\},
    \end{aligned}
\end{equation}
where
\begin{equation}
    W(k_1,k_2,m)=\sum_k\left[1+(-1)^k\right]\sqrt{\frac{(k-2)!}{(k+2)!}}(2k+1)\begin{pmatrix}
k_1 & k_2 & k \\
0 & 0 & 0 \\
\end{pmatrix}\begin{pmatrix}
k_1 & k_2 & k \\
-m & m+2 & -2 \\
\end{pmatrix}.
\label{W}
\end{equation}
The derivation of $P_y$ follows a similar process as that of $P_x$, and the results show that replacing $\sin[\phi_{n_1l_1m}-\phi_{n_2l_2(m+1)}]$ in the expression of $P_x$ with $-\cos[\phi_{n_1l_1m}-\phi_{n_2l_2(m+1)}]$ yields the expression of $P_y$.

\section{Consistency of the Galilean boosted wave function\label{C}}
To demonstrate that the wave function $\widetilde{\psi}_A(t,\mathbf{x})$ satisfies the SP equation, we first need to determine the random phase averaged potential associated with this new wave function. Let $\mathbf{x}^\prime=\mathbf{x}+\mathbf{v}t$, and the new potential can be obtained by solving the Poisson equation
\begin{equation}
    \nabla^{\prime 2}\widetilde{\Phi}(\mathbf{x})=\nabla^2\widetilde{\Phi}(\mathbf{x})=4\pi Gm_a\langle|\widetilde{\psi}_A(t,\mathbf{x})|^2\rangle=4\pi Gm_a\langle|\psi_A(t,\mathbf{x}^\prime)|^2\rangle.
\end{equation}
This leads to $\widetilde{\Phi}(\mathbf{x})=\Phi(\mathbf{x}^\prime)$. Our objective is to verify the following equation
\begin{equation}
    i\hbar\frac{\partial}{\partial t}\widetilde{\psi}_A(t,\mathbf{x})=-\frac{\hbar^2}{2m_a}\nabla^2\widetilde{\psi}_A(t,\mathbf{x})+m_a\widetilde{\Phi}(\mathbf{x})\widetilde{\psi}_A(t,\mathbf{x}).
\end{equation}
The left-hand side can be expressed as
\begin{equation}
    \begin{aligned}
       &\quad i\hbar\frac{\partial}{\partial t}\left[\psi_A(t,\mathbf{x}^\prime)e^{i(-m_a\mathbf{v}\cdot\mathbf{x}-\frac{1}{2}m_a\mathbf{v}^2t)/\hbar}\right]\\
       &=\left[i\hbar\frac{\partial}{\partial t}\psi_A(t,\mathbf{x}^\prime)+i\hbar\mathbf{v}\cdot\nabla^\prime\psi_A(t,\mathbf{x}^\prime)+\frac{1}{2}m_a\mathbf{v}^2\psi_A(t,\mathbf{x}^\prime)\right]e^{i(-m_a\mathbf{v}\cdot\mathbf{x}-\frac{1}{2}m_a\mathbf{v}^2t)/\hbar},
    \end{aligned}
\end{equation}
and the right-hand side as
\begin{equation}
    \begin{aligned}
        &\quad -\frac{\hbar^2}{2m_a}\nabla^2\left[\psi_A(t,\mathbf{x}^\prime)e^{i(-m_a\mathbf{v}\cdot\mathbf{x}-\frac{1}{2}m_a\mathbf{v}^2t)/\hbar}\right]+m_a\Phi(\mathbf{x}^\prime)\psi_A(t,\mathbf{x}^\prime)e^{i(-m_a\mathbf{v}\cdot\mathbf{x}-\frac{1}{2}m_a\mathbf{v}^2t)/\hbar}\\
        &=\left\{-\frac{\hbar^2}{2m_a}\left[\nabla^{\prime 2}\psi_A(t,\mathbf{x}^\prime)-2\frac{im_a\mathbf{v}}{\hbar}\cdot\nabla^\prime \psi_A(t,\mathbf{x}^\prime)-\frac{m_a^2\mathbf{v}^2}{\hbar^2}\psi_A(t,\mathbf{x}^\prime)\right]+m_a\Phi(\mathbf{x}^\prime)\psi_A(t,\mathbf{x}^\prime)\right\}e^{i(-m_a\mathbf{v}\cdot\mathbf{x}-\frac{1}{2}m_a\mathbf{v}^2t)/\hbar}.
    \end{aligned}
\end{equation}
Given that $\psi_A(t,\mathbf{x}^\prime)$ satisfies the Schr$\ddot{\text{o}}$dinger-Poisson equation, it follows naturally that the above two expressions are equal.
\end{widetext}
\end{appendix}
\end{document}